# SystemC Analysis of a new Dynamic Power Management Architecture

*Massimo Conti*
Università Politecnica delle Marche, via Brecce Bianche, I-60131, Ancona, Italy

**Abstract**

*This paper presents a new dynamic power management architecture of a System on Chip. The Power State Machine describing the status of the core follows the recommendations of the ACPI standard. The algorithm controls the power states of each block on the basis of battery status, chip temperature and a user defined task priority.*

## 1. Dynamic Power Management Architecture

Dynamic Power Management (DPM) is a design methodology that dynamically switches the operating mode of a system increasing or decreasing the performances of the system itself in order to reduce power consumption. Many DPM algorithms have been introduced to force in sleep or standby states the device when it is idle. Recently Intel, Microsoft and Toshiba proposed the Advanced Configuration and Power Interface (ACPI) to provide a standard for the HW/SW interface.

The dynamic power management architecture of a System on Chip (SoC) presented in this paper is reported in Fig.1. The SoC consists of different Intellectual Properties (IP). Each IP is considered as a black box and no detailed knowledge of its internal structure is assumed.

A Power State Machine (PSM) and a Local Energy Manager (LEM) are hardware components associated to each IP, or some of them.

The Power State Machine (PSM) describes the status of the core following the indications of the ACPI standard: executing at high or low performance, sleeping or in software-off state. The LEM dynamically switches the power state of the IP on the basis of the operations the IP is performing, and under the control of a Global Energy Manager (GEM), if it is present in the SoC.

The GEM conditionally enables the LEM on the basis of the requests of all the IP blocks, of the status of the SoC resources (battery energy, chip temperature, bus occupation, etc.), and on the priority of each IP. The DPM architecture proposed combines variable-voltage technique and the strategy that sets to sleep mode the device when it is inactive.

### 1.1. Functional IP

We suppose that the instructions, that the functional IP executes, are grouped in "tasks", that is sequences of instructions. The IP executes the "tasks" on the basis of some external service requests coming from the other IP blocks or from outside the SoC. The functional IP sends a task execution request to the LEM before the execution of each task. The LEM defines the power state of the PSM on the basis of GEM acknowledge, and the PSM enables the functional IP for the execution of the instruction according to the power state.

### 1.2. Power State Machine (PSM)

The PSM follows the recommendations of the ACPI standard: soft off, four sleep states ($SL_1$, $SL_2$, $SL_3$, $SL_4$), four execution states ($ON_1$, $ON_2$, $ON_3$, $ON_4$) with decreasing speed and power consumption using the variable-voltage technique. The voltage-scaling technique optimizes power consumption decreasing clock frequency and supply voltage in an appropriate way. During the power characterization of the IP an average energy dissipation is associated to each power state and type of instructions the IP is executing.

The DPM algorithm used considers the cost in terms of delay and power dissipation of the transition between two power states. The LEM sets the power state to the PSM that communicates the actual state to the functional block. The LEM estimates the consumption of the actual task on the basis of the signals coming from the PSM.

### 1.3. Local Energy Manager (LEM)

The Local Energy Manager (LEM) establishes the type of ON state when the IP must execute a task, and establishes if the IP must go in sleep or off state when it is inactive for a certain amount of time. The LEM receives in input the task priority (coded in 4 classes: Low, Medium, High and Very high), the battery status (coded in 5 classes: Empty, Low, Medium, High and Full), the chip temperature (coded in 3 classes: Low, Medium and High),

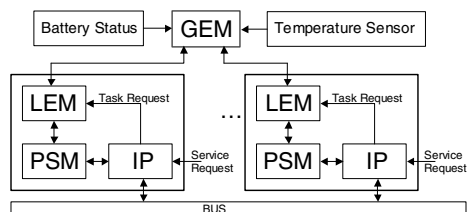

Fig. 1 - SoC dynamic power management architecture.

the energy dissipated by the other IP blocks in the SoC, and the enable signal from the GEM.

When the IP sends a task execution request, the LEM forwards the request to the GEM, if present, it estimates the battery status and temperature value at the end of the task execution and it establishes the power state following the rules reported in Table 1. These relationships can be seen as expressions of the natural language, as in the fuzzy rules:

```
If the priority is high and the battery is empty
   then the power state is ON4
…
```

These rules are a possible compromise between low energy consumption, low temperature and low latency.

The LEM dynamically switches the IP in sleep or off operating mode, when the IP is inactive. The manager makes a prediction of the idle time. This prediction is compared with the minimum time for which the state switching guarantees an reduction of energy dissipation, called break-even time.

### 1.4. Global Energy Manager (GEM)

The GEM receives resource requests from all the IP blocks of the SoC and it defines a static priority to each IP. The GEM returns to each LEM the energy requested by the other IP blocks, in such a way the LEM can correctly estimate the value of the battery status and chip temperature at the end of the task. The GEM can force each PSM in Sleep1 state if the resources are limited and the IP has low priority. The algorithm is the following

```
if (battery is Medium or High or Full)
    and (temperature is Low or Medium):
         enable every IP;
else if (battery is Empty or Low)
    and (temperature is Low or Medium):
         enable IPs with high priority;
else
    do not enable any IP
    switch on a supplementary fan
```

The GEM algorithm is intentionally simple. The complexity and the flexibility of the power management are left to the LEM, whose parameters can be adapted to the single IP to optimize its performances.

SystemC models of the battery and of the thermal sensor have been developed in order to verify the performances of the power management in different conditions.

## 2. SystemC Simulations and Results

The DPM architecture proposed has been implemented and simulated in SystemC 2.0. The variable-voltage technique has been simulated in SystemC defining the supply voltage as a variable that can change during simulation, this feature is not allowed in commercial power simulation tools. The functional IP blocks are pure traffic generators, but the application to real IP blocks is straightforward. Each IP executes a sequence of tasks or remains in idle state for a fixed time. Different types of input statistics have been considered in the simulations, in some sequences the IP is often busy , in some it is often in idle state.

Four SystemC simulations have been performed with just one LEM, PSM and IP executing the same sequence of tasks with different conditions:
A1) Battery Full and Temperature Low
A2) Battery Low and Temperature Low
A3) Battery Full and Temperature High
A4) Battery Low and Temperature High

Two SystemC simulations have been performed with a GEM and four LEMs, PSMs and IP blocks executing the following task sequences:
B) Battery Low and Temperature Low: IP1 with priority 1 (highest) and high activity, IP2 with priority 2 and high activity, IP3 with priority 3 and low activity, IP4 with priority 4 and low activity.
C) Battery Low and Temperature Low: IP1 with priority 1 and low activity, IP2 with priority 2 and low activity, IP3 with priority 3 and high activity, IP4 with priority 4 and high activity.

Table 2 reports the energy saving, temperature reduction and delay overhead with respect to the values required for the task execution at the maximum clock frequency without going to sleep or off mode.

In conclusion the DPM algorithm shows a high flexibility, allowing a bigger energy saving, but high delay, only when necessary. Furthermore the dependence of the power mode on the priority of each task allows an additional flexibility. The DPM algorithm is very efficient in the control of chip temperature. The simulation speed was 35 Kcycle/sec (sim. A) and 7.5 Kcycle/sec (B and C).

| Task priority | Battery | Temperature | Selected State |
|---|---|---|---|
| V | E | - | ON4 |
| V | - | H | ON4 |
| H, M, L | E | - | SL1 |
| H, M, L | - | H | SL1 |
| - | L | M, L | ON4 |
| - | E | M | ON4 |
| V | M, H | L | ON1 |
| H | M, H | L | ON2 |
| M | M, H | L | ON3 |
| L | M, H | L | ON4 |
| V, H, M | F | L | ON1 |
| L | F | L | ON2 |
| - | Power supply | M, L | ON1 |

Table 1 – Power state selection algorithm

|    | Energy saving (%) | Temperature reduction (%) | Average Delay overhead (%) |
|---|---|---|---|
| A1 | 39 | 31 | 30 |
| A2 | 55 | 21 | 339 |
| A3 | 39 | 18 | 37 |
| A4 | 55 | 18 | 339 |
| B  | 65 | 19 | 242 |
| C  | 64 | 18 | 253 |

Table 2 – Performances of the DPM in the different simulations